# Significant Photoluminescence Improvements from Bulk Germanium-Based Thin Films with Ultra-low Threading Dislocation Densities

Liming Wang[1], Gideon Kassa[2], Aofeng Bai[2], Jifeng Liu[2], and Guangrui Xia[1]*

[1] Department of Materials Engineering, The University of British Columbia, Vancouver, BC V6T 1Z4, Canada

[2] Thayer School of Engineering, Dartmouth College, Hanover, New Hampshire 03766, United States

*Corresponding author: guangrui.xia@ubc.ca

**Abstract**

Bulk Ge crystals, characterized by significantly lower threading dislocation densities (TDD) than their epitaxial counterparts, emerge as optimal candidates for studying and improving Ge laser performance. Our study focused on the Ge thickness and TDD impacts on Ge's photoluminescence (PL). The PL peak intensity of a bulk Ge sample (TDD = 6000 $cm^{-2}$, n-doping = $10^{16}$ $cm^{-3}$) experiences a remarkable 32-fold increase as the thickness is reduced from 535 μm to 2 μm. This surpasses the PL peak intensity of a best-performing epitaxial-Ge on Si (epi-Ge) (0.75 μm thick, biaxial tensile strain= 0.2%, n-doping = 7 ×$10^{18}$ $cm^{-3}$) by a factor of 2.5. Furthermore, the PL peak intensity of a 405 μm thick zero-TDD bulk Ge sample (n-doping = $2.5 \times 10^{18}$ $cm^{-3}$) is 9.7 times that of the 0.75 μm thick epi-Ge, rising to 12.1 times when thinned to 1 μm. Although the bulk Ge-based TDD reduction approach doesn't boost the direct band transitions, it can work alongside n-type doping and strain engineering to enhance Ge laser performance and relax the requirement on the latter two approaches, which reduce the associated side effects of high optical absorption, high non-radiative recombination, and large footprint.

# 1. Introduction

Silicon (Si)-compatible on-chip lasers are important across diverse applications such as optical computing.[1], bio-chemical sensing[2], quantum computing[3], and on-chip and off-chip optical communication[4]. [5] Over the past few decades, researchers worldwide have dedicated extensive efforts to identifying viable solutions for a Si-compatible light-emitting material system.[6-8] A promising solution in addressing this challenge is the adoption of germanium (Ge), recognized for its exceptional compatibility with Si, making it a well-suited semiconductor for monolithic integration. Furthermore, Ge has already found applications in MOSFETs[9-10] and Si photonics, serving as detectors[11-12] and modulators[13-14]. However, like silicon, Ge is an indirect bandgap material, generally considered a poor light-emitting material.

The indirect bandgap of Ge is 0.664 eV at the L valleys, and the direct bandgap is 0.800 eV at the $\Gamma$ valley, exhibiting a difference of 136 meV at room temperature.[15] Ge can be transformed into a direct or pseudo-direct bandgap material through bandgap engineering by adding tensile strains and/or n-doping, providing a key advancement that facilitates efficient light emission.[16] A milestone in developing Ge on-chip lasers was achieved in 2012 by demonstrating the first electrically pumped Ge laser with 0.2% biaxial tensile strain and $4 \times 10^{19}$ cm$^{-3}$ n-type doping.[17] However, initial Ge lasers faced challenges, including high threshold currents of approximately 280 kA/cm² and low efficiencies ranging from 0.5% to 4%, thus presenting obstacles for practical applications. One of the reasons for such poor performance can be attributed to the high TDD in the $10^6$ to $10^{10}$ cm$^{-2}$ range in epitaxial Ge (epi-Ge) on Si.[18] In this paper, all epi-Ge films refer to epi-Ge films on Si. TDDs contribute to non-radiative recombination processes, leading to higher threshold currents and reduced efficiencies.[19] In contrast, bulk Ge crystals have ultra-low TDD of $\leq 10^4$ cm$^{-2}$, and the best commercial Ge wafers are TDD-free when measured using the etch pit density (EPD) method. The ultra-low TDDs make bulk Ge valuable in exploring the ultimate potentials of Ge on-chip lasers and Ge laser improvement strategies.

While making a Ge laser requires takes solving many process challenges due to the much worse manufacturability compared to Si, photoluminescence (PL) measurements are widely used to study radiative and non-radiative recombination in Ge in the early stages.[20-21] PL properties such as PL peak intensities, shapes, integrated areas, and peak wavelengths are important measurements of the relevant bands for light emissions and emission potentials. Owing to the indirect band structure, Ge's $\Gamma$ and L valleys should generate two peaks in PL. Most PL studies focus on epi-Ge, which revealed that the PL intensity can be improved with larger tensile strain.[22-23], higher n-doping[21, 24-25], higher temperature[24], and a larger laser power[26], the PL results of bulk Ge from different literature can contradict each other. For example, Haynes et al. reported that room temperature PL from bulk Ge could only be observed when self-absorption was reduced[27]. More study seems to support this conclusion as PL from bulk Ge could only be detected at low temperatures.[28]. However, Jan et al. demonstrated that an n-doped bulk Ge wafer (500 μm thick, n-doping =1 x 10$^{15}$ cm$^{-3}$) had 15 times the integrated intensity of PL of epitaxial Ge-on-Si (2.8 μm thick, n-doping < 1 x 10$^{17}$ cm$^{-3}$, EPD = 3.7 x 10$^6$ cm$^{-2}$) measured at room-temperature due to a much lower dislocation density.[29] To avoid the n-doping introduced bandgap engineering, the epi-Ge sample in that study was unintentionally doped, which doesn't compare with highly n-doped best-performing epi-Ge films. The development of bulk Ge-related PL studies is summarized in Table 1.

**Table 1**. A summary of Bulk Ge PL studies from the literature

| Ge type | Thickness | Measurement temperature | Major conclusions | Reference | Limitations |
|---|---|---|---|---|---|
| Bulk Ge | ≤ 10 μm | Room temperature | Direct gap PL is observed when bulk Ge is thinned down to less than 10 μm | 27 | No comparison between epi-Ge and bulk Ge |
| Bulk Ge | 100 μm | 295 K,174 K | (i) Peak shift from1.76 μm to 1.54 μm with increased power (ii) Direct band transition is more efficient | 30 | No comparison between epi-Ge and bulk Ge |
| Bulk Ge GeOI | - | 300 K | 20 X enhancement compared to the undoped material near the 1550 nm for active dopant concentrations around 5×10$^{19}$ cm$^{-3}$ | 21 | No comparison between epi-Ge and bulk Ge |
| Bulk Ge Epi-Ge | - | Room temperature | (i) Indirect band emission dominates in Bulk Ge; | 31 | No direct comparison between epi-Ge and bulk |
| (n doping = $1.9 \times 10^{19}$ cm$^{-3}$) | | | direct band emission dominates in epi-Ge (ii) This is attributed to a lack of self-absorption in thin Ge films | | Ge, bulk Ge data from the literature |
| Bulk Ge Epi-Ge (n doping = $1 \times 10^{17}$ cm$^{-3}$) | | Room temperature | (i) Bulk Ge has 15 times the integrated intensity of photoluminescence of Ge-on-Si. (ii) Defects in the Ge-on-Si are responsible for the weak indirect transition and relatively strong direct transition | [29] | The doping concentration of epi-Ge is low, not the best epi |
| Bulk Ge | | 7 K | PL on intrinsic and doped bulk Ge substrates as a function of temperature and excitation power | [28] | Low-temperature PL, no comparison between epi-Ge and bulk Ge |
| Bulk Ge Ge microstrip | | 100 K-375 K | Increasing PL intensity for rising values of strain, excitation power, and temperature | [26] | TDD or doping data are not available |
| Bulk Ge Epi-Ge (n doping = $4 \times 10^{18}$ cm$^{-3}$) | | Room temperature | The intensity ratio between the direct and indirect optical transition drastically decreases with decreasing temperature in both n-type epitaxial and p-type bulk Ge | [32] | p-doped bulk Ge demonstrated similar intensity to n-epi, with no comparison between n bulk vs. n epi |

Is bulk Ge able to surpass the best-performing n-doped epi-Ge films? What are the roles of TTD and thickness in determining Ge PL properties? Can the TDD reduction method be added to the existing strain and doping engineering to improve Ge's light emission? This work addressed these questions by investigating the PL properties of a bench-marking epi-Ge control sample with very high PL, two bulk Ge wafers and bulk Ge-based thin films with different TDD, doping, and thicknesses.

## 2. Experiment

### 2.1 Sample selection

The specific bench-marking epi-Ge sample was chosen as the control sample for this study. Detailed information about this sample has been previously documented in Ref. 31, where it was named P-NA-off. This bench-marking epi-Ge has a 5% higher PL intensity than that of the MIT Ge sample in Ref. 31, which is equivalent to the Ge samples that achieved the first and best performance to date for room temperature operated electrically pumped Ge lasers.[33] Therefore, this epi-Ge control sample represented the best-performing epi-Ge samples. Two Ge wafers from two sources were selected as the starting materials. The first was a 4-inch n-type Ge wafer purchased from University Wafer, Inc. (UW) with a typical TDD of 6000 cm$^{-2}$, as previously measured and reported in Ref 30.[34] The second wafer was a 4-inch n-type (with a resistivity of $\leq 0.01$ Ohms*cm at 295 K), single-side polished (100) Ge wafer sourced from Umicore N. V. in Belgium. Notably, this wafer has zero TDD measured by etch-pit density (EPD). The corresponding doping is $2.5 \times 10^{18}$ cm$^{-3}$; therefore, it is named Umicore18. The third wafer is $2.5 \times 10^{14}$ doped from Umicore and is named Umicore14. All the information on the bulk Ge wafers and the epi-Ge control is summarized in Table 2. In the following discussions, when the comparison is made among samples with different thicknesses, the Ge samples are named with the wafer name and the corresponding thickness such as "Umicore18-405 μm" and "UW-2 μm".

**Table 2**. The information on bulk Ge from UW and Umicore and the epi-Ge control sample. The TDD of the epi-Ge was measured by electron channeling contrast imaging (ECCI), which is a much more accurate method for measuring TDD than the EPD method for doped Ge.

| Ge type and sample notations | | Thickness (μm) | Tensile strain | P concentration (cm$^{-3}$) | S$_q$ (nm) | TDD (cm$^{-2}$) |
|---|---|---|---|---|---|---|
| Epi-Ge | The epi-Ge control | 0.75 | 0.2% (biaxial) | $7 \times 10^{18}$ | 0.8 | $5.2 \times 10^8$ |
| Bulk Ge wafers | UW | 535 | 0 | $1 \times 10^{16}$ | 1.6 | $\sim 6000$ |
| | Umicore18 | 405 | 0 | $2.5 \times 10^{18}$ | 1.0 (polished side) 8.2 (unpolished side) | 0 |
| | Umicore14 | 225 | 0 | $2.5 \times 10^{14}$ | 0.5 | 0 |

## 2.2 Bulk Ge thinning with wet etching

The UW wafers and Umicore18 wafers were cut into small pieces before the wet etching. HCl-$H_2O_2$-$H_2O$ solutions with a volume ratio of 1:1:5 were chosen for the wet etching according to our previous wet etching study that produced 535-μm-thick Ge thin films down to 4.1 μm with good integrity and no increase in the TDD.[34] The thinning process was the same as we have reported in Ref 35.[35] The details of wet etching are included in Supporting Information.

## 2.3 Thin Film Characterizations

The thickness of the Ge thin film was measured using a Nikon ECLIPSE LV150 optical microscope equipped with length measurement capabilities. Simultaneously, 3D optical profiler images and rough mean square surface roughness ($S_q$) of the Ge samples were measured using an optical interferometer (Filmetrics Profilm3D optical surface profiler). Three different positions on each sample were checked both for the thickness and the surface roughness measurement.

To explore the PL properties of Ge, a commercial Horiba LabRAM HR Evolution instrument was employed, featuring an InGaAs photodetector cooled with liquid nitrogen. In this setup, a 1064 nm laser was applied to all samples, a 300 gr/mm grating was used, and a 50X objective lens was utilized. The spot size was around 1 μm, and the acquisition time for each spot was set at 3 seconds. Different filters were used to attenuate the excitation laser power before it entered the sample, which translated to 5, 10, and 25% of the original laser power of about 20 mW on the sample surface without any filtering. The choice of laser power was to achieve a balance between a good signal-noise ratio and mitigating the heating effect. The localized heating on the sample may indeed facilitate increased electron injection into the conduction band, resulting in higher PL intensity and a redshift in the PL peak. However, it's crucial to minimize the heating effect to accurately assess the true PL of the sample, without interference from heating-induced changes.

A 10% laser power was employed for most measurements with a satisfactory balance between the signal strength and minimal heating. By comparing the PL peak positions, we observed that the PL peaks with 5% and 10% laser power were similar at 1536 nm, but both were smaller than those with 25% laser power by 22 nm as seen in Fig. 4 (c). This indicated that the heating effect from 10% laser power was negligible. In the figures below, the default measurement laser power condition was 10% laser power unless otherwise noted. In cases where UW bulk Ge failed to yield sufficient signal at 10%, a 25% laser power was utilized. The PL spectra were first smoothed and then deconvoluted into two peaks. The peak with a shorter wavelength (close to 1550 nm for bulk Ge and bulk Ge-based thin films) was attributed to the direct band transition, and another peak with a longer wavelength was assigned to the indirect band transition.

### 3. Results and discussion

### 3.1 PL from UW bulk Ge and thinned UW Ge

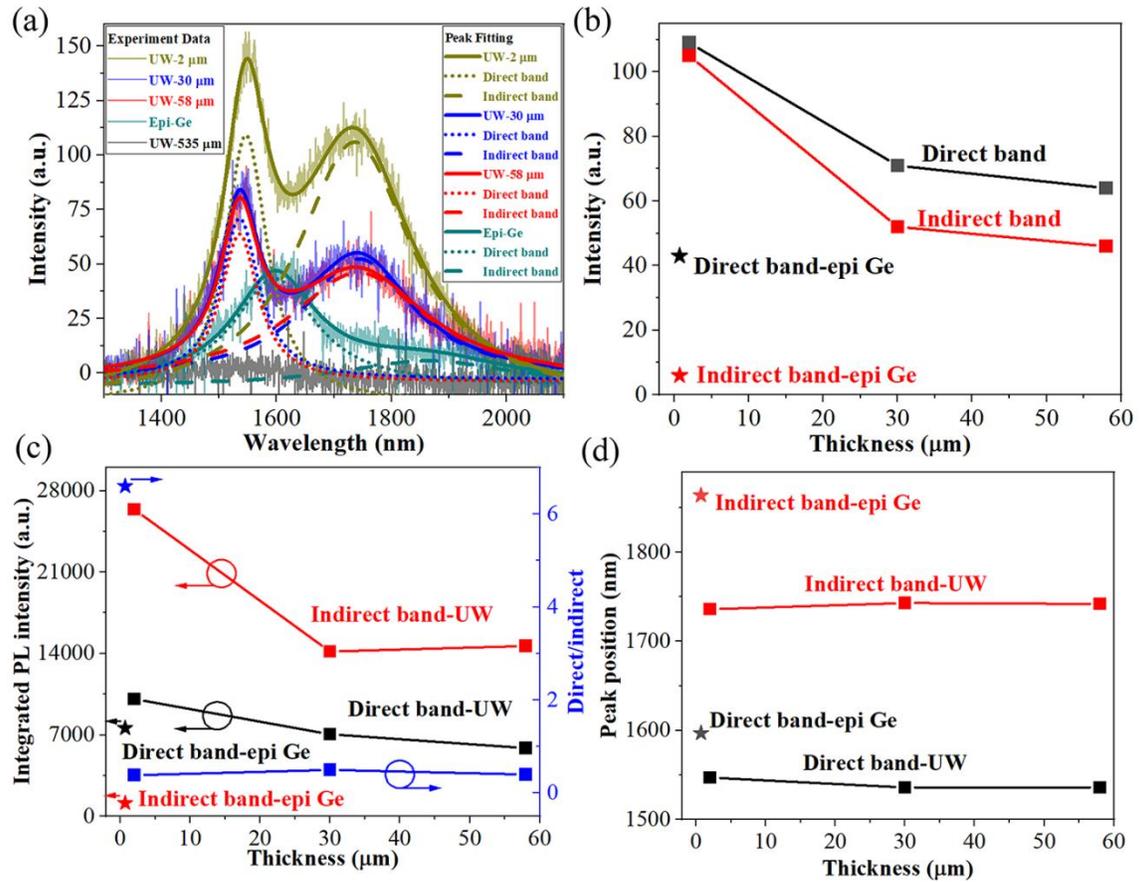

**Figure 1.** PL measurements were conducted with 10% of the laser power. (a) the measured PL from the epi-Ge control and UW Ge with different thicknesses, and the fitted PL peaks from the epi-Ge control and UW Ge with different thicknesses (except for UW Ge-535 μm), (fitted PL peaks are shown except for UW Ge-535 μm), (b) the PL intensity (peak height) vs. UW Ge thickness,(c) the integrated PL intensity (peak area) vs UW Ge thickness, and the ratio of the integrated PL intensity of the direct peak over that of the indirect peak, and (d) the PL peak wavelength position vs. UW Ge thickness.

Figure 1 compares the UW Ge samples and the epi-Ge control sample. The epi-Ge used in this experiment is comparable to the PL achieved in the previously reported MIT sample[33], which exhibits two distinct PL peaks after peak finding and fitting (Figure 1a). One represents the direct band emission at approximately 1597 nm, and the other corresponds to the indirect band emission at around 1864 nm. The red-shifted direct band emission at 1597 nm (Figure 1d), in contrast to the expected 1550 nm from relaxed Ge, is attributed to a biaxial tensile strain of 0.2% and heavy n-doping of $7 \times 10^{18}$ cm$^{-3}$ in the epi-Ge[33].

Notably, the thickness of UW Ge emerges as a crucial factor influencing the PL spectra. The PL of UW-535 μm is obscured with a substantial thickness due to self-absorption from the indirect band.[34] Reducing the thickness to less than 60 μm diminishes the self-absorption, significantly increasing the PL signal. A comparison between UW-58 μm and the epi-Ge control reveals that, despite the epi-Ge control having 700 times higher P doping concentration and 0.2% tensile strain, the direct peak intensity from UW-58 μm surpasses that of the epi-Ge control (Figure 1a and 1b). The PL intensity increases more when the thickness is reduced to 30 μm. When the thickness decreases from 58 to 2 μm, the peak intensity (peak height) from the direct band is 1.7 times that of the UW-58 μm, 2.5 times higher than the epi-Ge (Figure 1b). This unexpected outcome underscores the prominence of TDD in improving PL intensities.

As shown in Figure 1c, the integrated PL intensity from the direct band of UW samples also keeps increasing with thinner thicknesses. The integrated intensity of the direct band increases by 1.7 times when the thickness decreases from 58 μm to 2 μm. Even though the direct peak intensity of the UW sample is higher than that of the indirect peak, the integrated intensity of the indirect peak is significantly greater. This can be attributed to the smaller bandgap of the indirect band compared to the direct band, making recombination at the indirect band more energetically favorable. Also, in these free-standing Ge thin films, no tensile strains were added to boost direct band transitions. The ratio of the integrated direct band PL peak to that of the indirect band PL peak in the UW samples remains around 0.4 to 0.5 for the thickness ranges from 2 to 58 μm, which is significantly lower than the ratio 6.6 of the epi-Ge control. This indicates that while reducing TDD enhances PL intensity by reducing non-radioactive recombination centers, it doesn't help to promote direct band transitions, as the bandgap engineering techniques, i.e., tensile strain introduction and n-type doping.

The peak positions are summarized in Figure 1d, the associated direct peak position of the UW-58 μm sample is approximately 1536 nm, closely aligning with literature reports at 1540 nm.[21] This alignment suggests that using 10% of the laser power is suitable for minimizing heating effects during PL measurements. The direct peak position remains similar for the UW-30 μm sample but slightly increases when thinned to 2 μm due to unavoidable heating effects, which are more pronounced for thinner films. The indirect peak position remains around 1745 nm, corresponding to an energy of 0.71 eV, which is significantly higher than the well-accepted 0.66 eV for the Ge bandgap. This discrepancy is attributed to the phonon involvement in the indirect band transition, where phonon energy is absorbed during the transition process. [36]

### 3.2 PL of the UW Ge with a higher excitation laser power

As discussed in the introduction, observing PL from bulk Ge can be difficult. Since no obvious peak can be observed for UW-535 μm bulk Ge with a 10% laser power, a higher excitation laser power is applied for the UW bulk Ge. For UW-535 μm, 25% laser power ensures an adequate PL signal (Figure 2) over the background noise. Under this condition, compared to UW-535 μm, the UW-2 μm exhibits a remarkable 32-fold increase in the PL intensity, emphasizing a significant enhancement in PL intensity with reduced thickness.

It's crucial to note that the excitation laser heating effect plays a pivotal role in the thinner regions. With a higher temperature (power), more electrons are pumped to the direct valley, which is evident in the dramatic increase in direct emission, and a decrease in indirect emission, which is well aligned with the feature of the direct band and the indirect band transitions. A peak shift from 1556 nm to 1594 nm is observed as the thickness reduces from 535 μm to 2 μm indicating obvious heating with larger laser power. If the heating effect can be removed for this measurement condition without changing the input laser power, the PL peak intensity of UW-2 µm is expected to be more than 32X that of UW-535 µm.

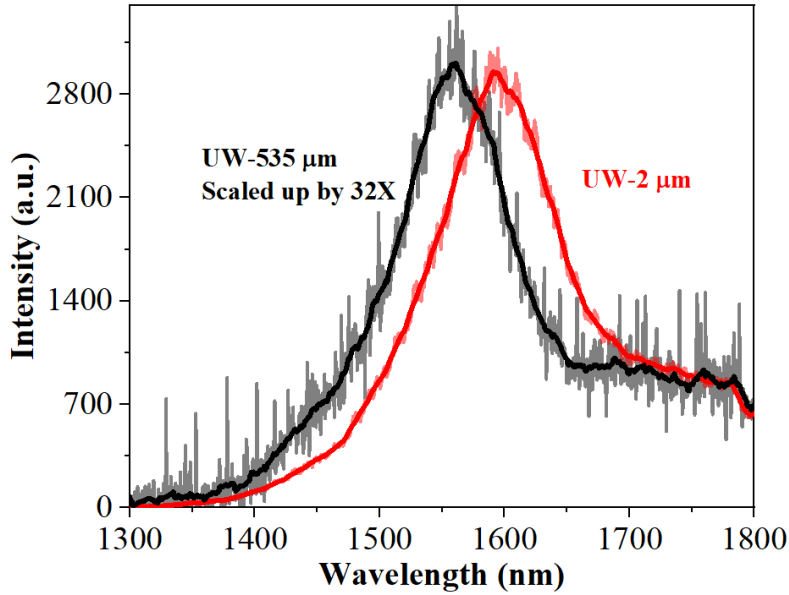

**Figure 2.** The PL spectra of UW-535 µm scaled up by 32X and UW-2 µm measured with a 25% laser power. The scaling is to make the UW-535 µm spectrum easier to read.

### 3.3 PL of Umicore18 Bulk Ge (ultra-low TDD)

The PL intensity from Umicore18 bulk Ge is significantly higher than that of UW bulk Ge, 9.7 times that of epi-Ge control (Figure 3). This difference may explain the discrepancy in bulk Ge PL measurements discussed in the previous literature. The variation in results can be attributed to the TDD within the bulk Ge. When the TDD is high, PL from bulk Ge is barely detectable. In cases where the TDD is in a medium range, such as in UW Ge, PL from bulk Ge can be observed but requires higher excitation power. For Umicore18 Ge, the zero TDD allows the PL signal to be easily observed even with a substantial thickness.

The PL intensity consistently increases with decreasing thickness, resembling the trend observed in UW Ge. Remarkably, after thinning the bulk Ge to 1 µm (near the top), the intensity increases to 12.1 times that of the epi-Ge control (Figure 3c). Notably, the thickness effect on Umicore18 bulk

Ge is less pronounced than the UW Wafer. The intensity for Umicore18 Ge increases by only 24% when the thickness is reduced from 405 μm to 1 μm, whereas the intensity of UW Ge increases nearly 32 times when the thickness decreases from 535 μm to 2 μm. This suggests that the self-absorption is much stronger for high TDD Ge films, whose PL spectra have much stronger thickness dependence.

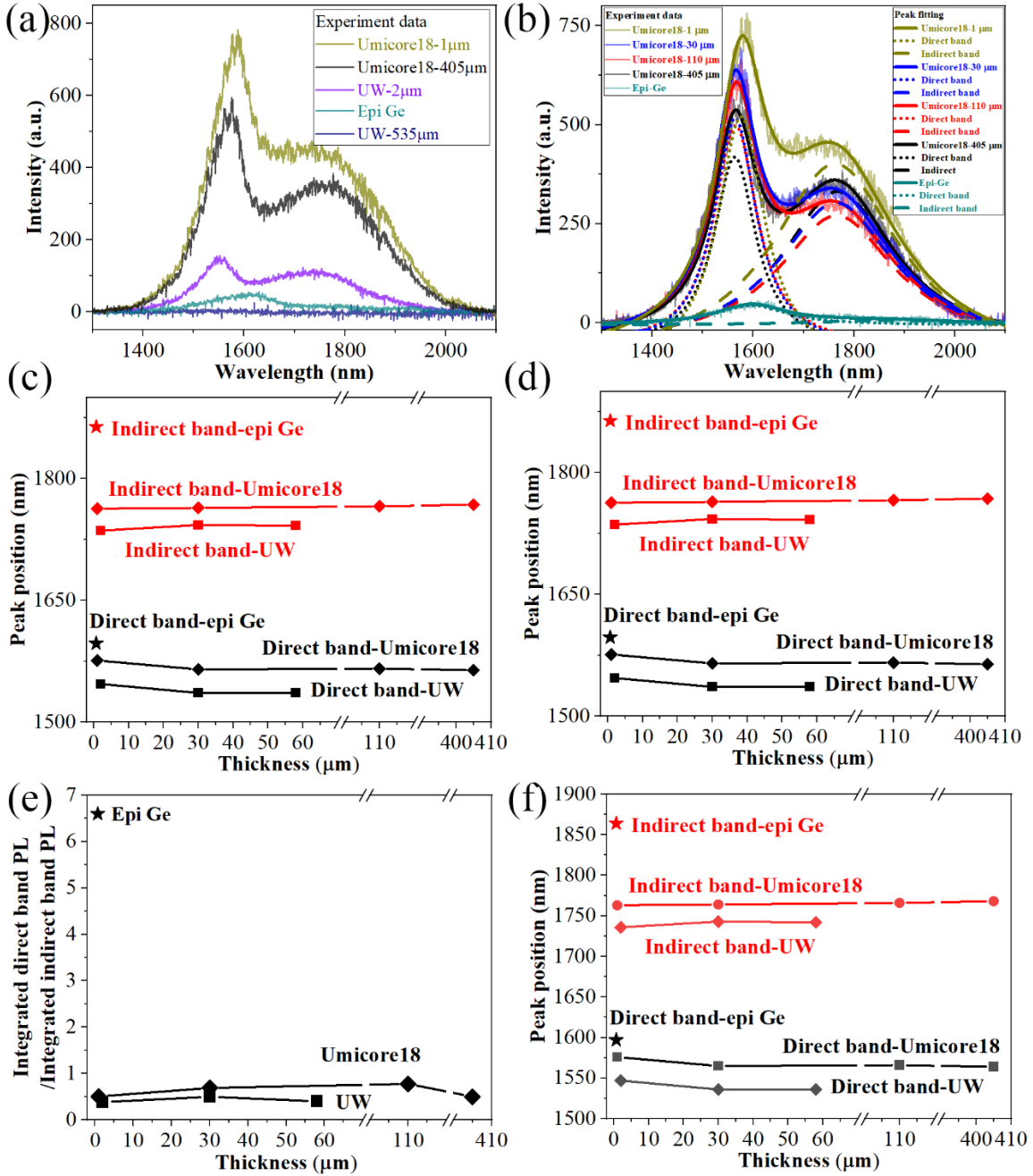

**Figure 3.** (a) The PL of the Umicore18 bulk Ge and Umicore18 Ge thin film with different thicknesses, with a 10% laser power (b) the fitted PL peaks of the Umicore18 bulk Ge and Umicore18 Ge thin films with different thicknesses, (c) the PL intensity vs Ge thickness, (d) the integrated PL intensity vs Ge thickness, (e) the ratio of the integrated area of the direct peak over the indirect peak, and (f) the PL peak position vs Ge thickness.

## 3.4 PL excitation laser power dependence of the Umicore18 Ge PL

It has been demonstrated that higher laser power heats the sample surface, injecting more electrons into the conduction band and thereby increasing the PL intensity. The laser power dependence near the top of the Umicore18 thin film is shown in Figure 4a. It is noted that the integrated PL intensity increases significantly with higher input laser power. Meanwhile, the direct-to-indirect ratio also increases with more power, as more electrons are pumped into the direct band, enhancing the chances for direct band emission. Naturally, higher power causes the peak position to shift to a longer wavelength. Therefore, selecting the optimal laser power is crucial for obtaining reliable PL results.

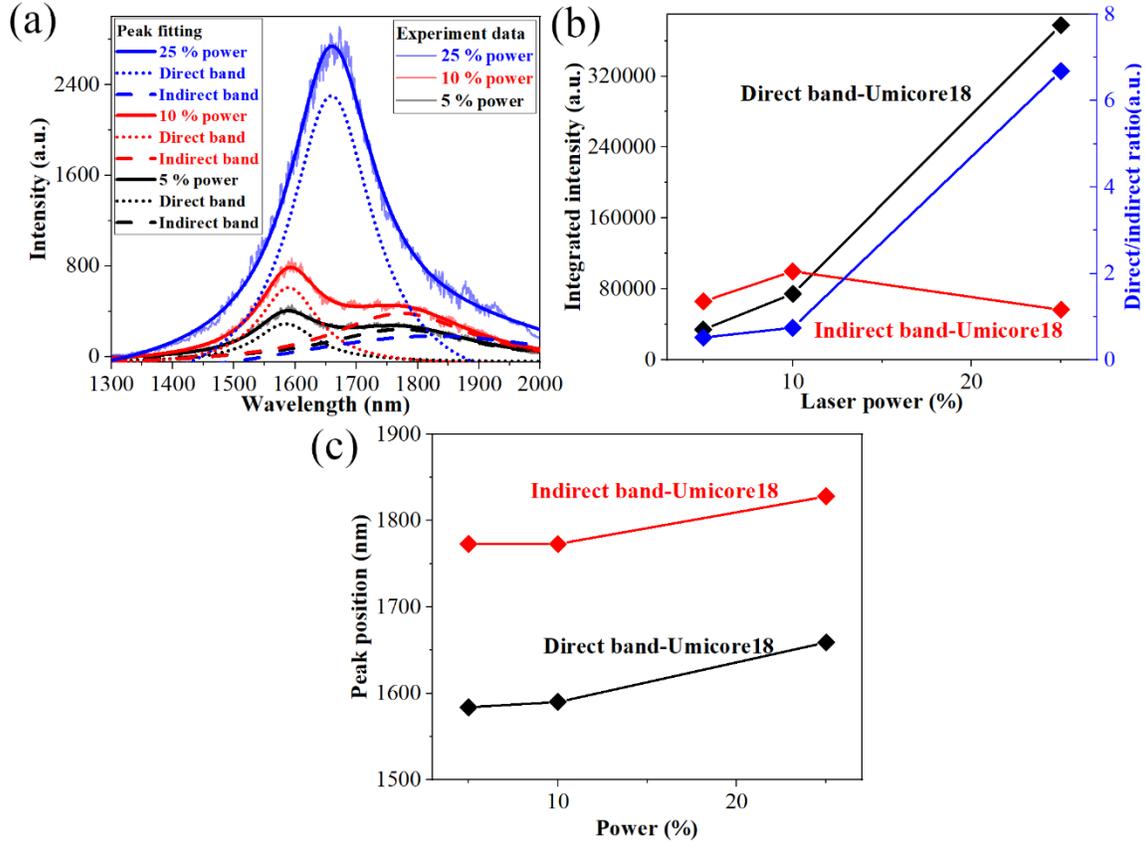

**Figure 4.** (a) The laser power dependence near the edge of the Umicore18 thin film, (b) the PL intensity vs. laser power, (c) the PL peak position vs laser power.

### 3.5 N doping dependence of the Ge PL

Even though Umicore18 bulk Ge exhibited significantly higher PL intensity compared to UW bulk Ge and epi-Ge, a key question remains: how much of this enhancement is attributed to the lower TDD and how much is from the higher n-type doping? To disentangle these two factors, a $2.5 \times 10^{14}$ n-doped Umicore bulk Ge sample was used as a control (Umicore14). The PL results for different bulk Ge samples are presented in Figure 5, and the related information and PL intensity are summarized in Table 3. Comparing the UW-535 μm sample to the Umicore14, the latter showed significantly higher PL intensity, emphasizing the critical role of TDD. Furthermore, a comparison between the Umicore18-405 μm sample and the Umicore14-225 μm confirms that n-type doping enhances PL intensity. Notably, the Umicore18-405 μm sample and the Umicore14-225 μm exhibited higher PL intensity than epi-Ge, even with much lower doping concentrations, highlighting that TDD plays a more significant role than n-type doping in enhancing Ge PL intensity.

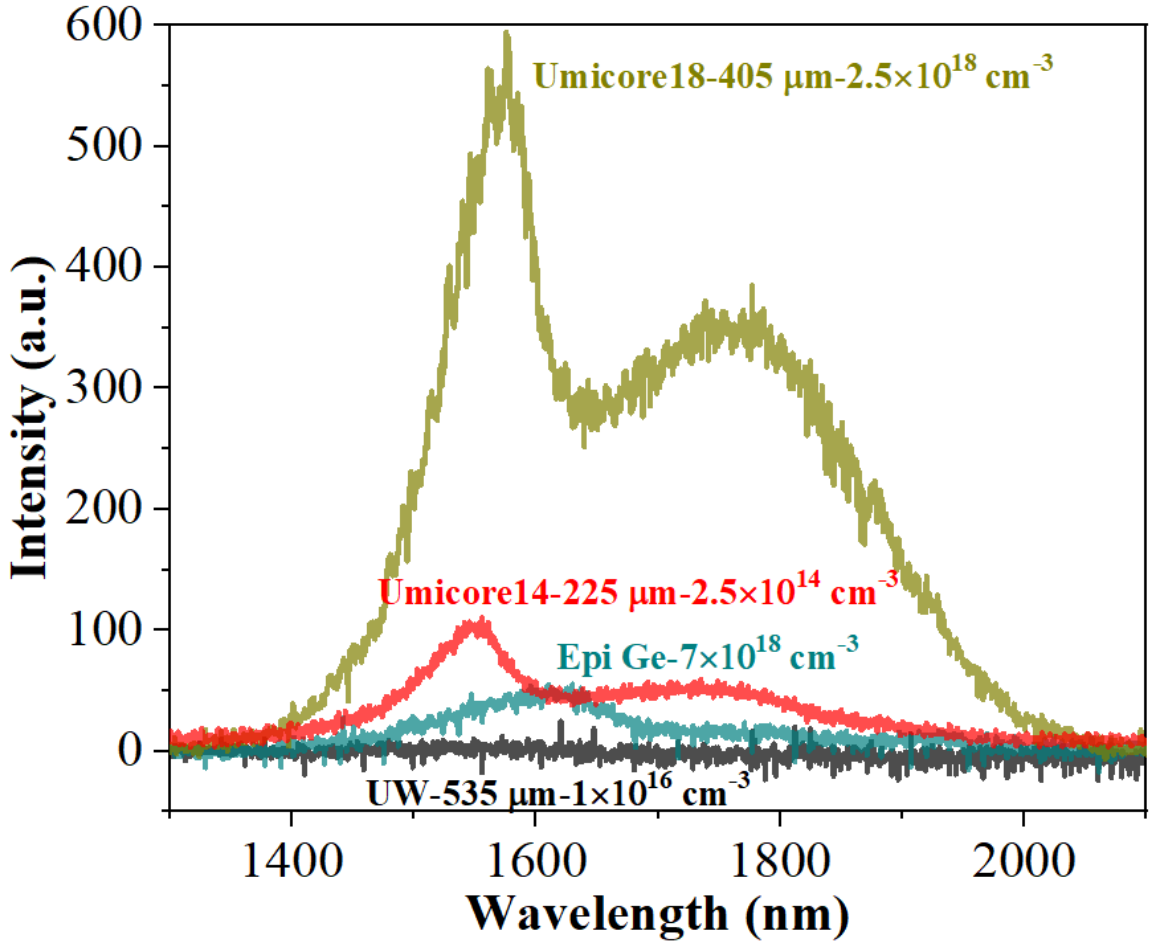

**Figure 5.** The PL spectra of UW-535 µm, epi-Ge, Umicore14, and Umicore-405 µm

**Table 3**. The information and PL intensity on bulk Ge from UW, Umicore, and the epi-Ge control sample, PL intensity higher than epi-Ge highlighted in green.

|  | Thickness (µm) | Tensile strain (%) | P concentration (cm$^{-3}$) | Surface roughness (nm) | TDD (cm$^{-2}$) | Relative PL peak intensity |
|---|---|---|---|---|---|---|
| **Benchmarking epi-Ge on Si** | 0.75 | 0.2 (biaxial) | $7\times10^{18}$ | 0.8 | $1\times10^{8}$ | 1 |
| **Original UW Ge** | 535 | 0 | $1\times10^{16}$ | 1.8 | $\sim 6000$ | $\sim 0$ |
| **Thinned UW Ge** | 2 | 0 | $1\times10^{16}$ | 10 | ~ 6000 | 2.5 |
| **Original Umicore18** | 405 | 0 | $2.5\times10^{18}$ | 1.0 | 0 | 10 |
| **Thinned Umicore18** | 1 | 0 | $2.5\times10^{18}$ | 1.9 | 0 | 12 |
| **Umicore14** | 225 | 0 | $2.5\times10^{14}$ | 0.5 | 0 | 2.1 |

### 3.6 Surface roughness dependence of the Umicore18 Ge film PL

A key consideration in wet etching is understanding whether surface roughness plays a role in determining PL intensity. To investigate this, we compared the Umicore18-110 μm on both sides, as depicted in Figure 6. Surprisingly, there is no noticeable difference in peak shape, indicating that surface roughness does not significantly influence it. Furthermore, there is no discernible difference in peak position, suggesting that surface roughness does not contribute to the peak position. The difference is subtle, while the PL intensity shows a minor increase (~ 5%) on the front side with lower surface roughness (Figure 8b). This observation suggests that surface roughness can play a role in PL intensity but is a minor contributing factor. Despite the significant difference in surface roughness, the impact on PL intensity remains relatively small.

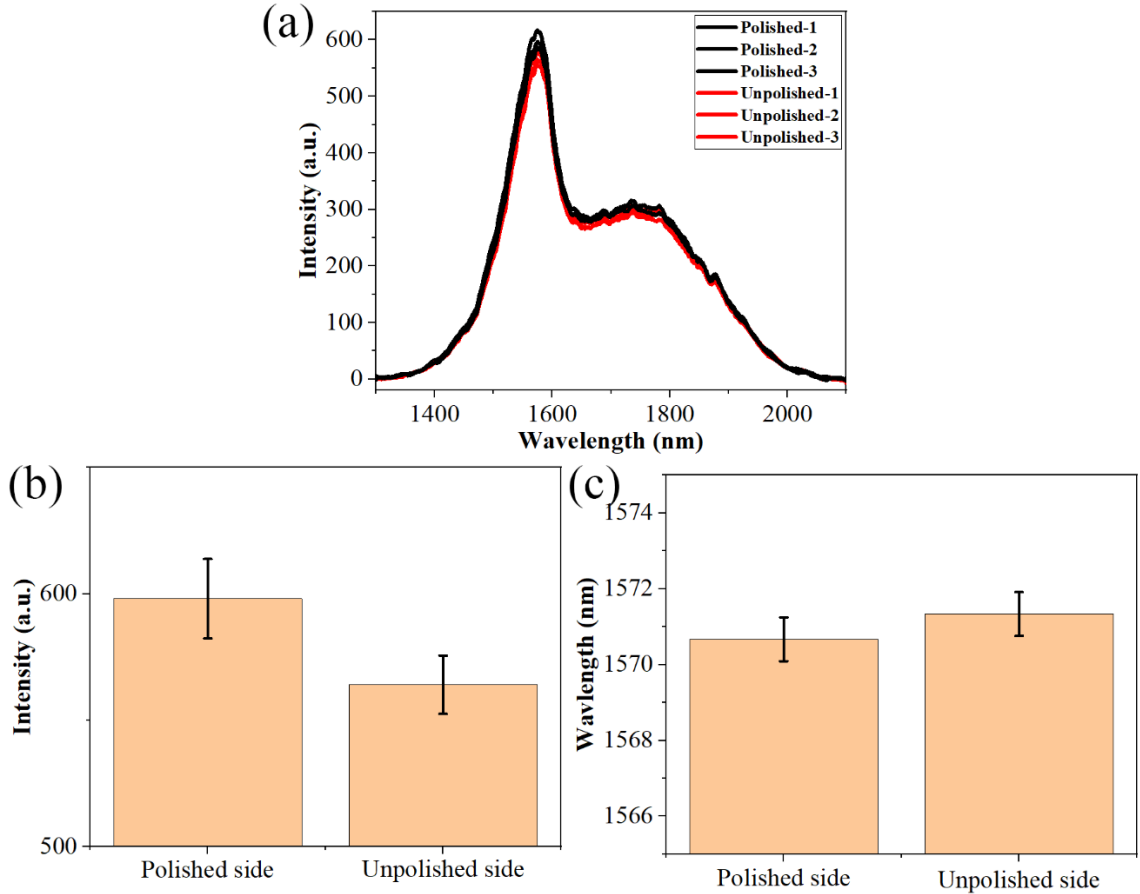

**Figure 6.** (a) The PL from both sides of the Umicore18 Ge-110 μm ($S_q$ of the polished side = 1.9 ± 0.3 nm, $S_q$ of the unpolished side = 7.4 ± 2.1 nm), (b) the peak intensity from both sides, (c) the peak position from both sides.

## 4. Conclusions

In this study, we demonstrated that the PL peak intensity of a UW bulk Ge sample (535 μm thick, TDD = 6000 cm$^{-2}$, n-doping = 1 × 10$^{16}$ cm$^{-3}$) can be significantly enhanced by a factor of 32 by reducing the thickness to 2 μm by wet etching. Capitalizing on the superior Ge quality, even with a 1.25 μm more thickness, no strain, and 1/700 of the n-doping, the PL peak intensity of the 2-μm-thick UW bulk Ge sample, surpasses that of the bench-marking epi-Ge sample (0.75 μm thick, TDD = 5.2 × 10$^8$ cm$^{-3}$, biaxial tensile strain= 0.2%, n-doping = 7 × 10$^{18}$ cm$^{-3}$) by a factor of 2.5, demonstrating the critical role of TDD reduction in PL intensity enhancement. Furthermore, our ultra-low-TDD Umicore18 Ge thin film (1 μm thick, TDD = 0, n-doping = 2.5 × 10$^{18}$ cm$^{-3}$) achieved 12.1 times the PL intensity of the bench-marking epi-Ge with 36% of the n-type doping. The increase in surface roughness from 1.9 to 7.4 nm has a negligible impact on the PL intensity.

Lower TDDs help to enhance the direct band PL and the indirect band PL by reducing the non-radiative recombination centers. Significant PL intensity enhancement was achieved without tensile strain or high n-type doping. The latter two approaches can be added to the TDD reduction approach readily. The TDD reduction approach relaxes the requirement of high n-doping and stress concentration, which also mitigates the side effects of high optical absorption, high recombination rates, and bandgap narrowing associated with the high n-type doping method and the larger footprint and bandgap narrowing associated with the stress concentration method. The use of ultra-low-TDD bulk Ge wafers as the starting materials was proved to overcome the technology barrier of high-TDDs in epi-Ge, and provides a third approach to improve on-chip Ge lasers.

**Author contributions**

Liming Wang: conceptualization, methodology, investigation, writing original draft, writing-review & editing, visualization. Gideon Kassa: PL investigation. Aofeng Bai: PL investigation. Jifeng Liu: research supervision in PL measurements, resources, writing- review & editing. Guangrui (Maggie) Xia: conceptualization, research supervision, resources, writing – review & editing.

**Conflicts of interest**

There are no conflicts to declare.

**Acknowledgement**

We acknowledge Umicore N.V. for their support in providing coated Ge wafers. CMC Microsystems is acknowledged forproviding MNT Awards to cover the fabrication costs. Our thanks extend to UBC Nanofab for their training and assistance with cleanroom equipment. The authors acknowledge UBC and the Natural Sciences and Engineering Research Council of Canada (NSERC) for funding this research. Additionally, Liming Wang acknowledges the financial support from UBC's Four Year Doctoral Fellowship.